\documentclass[reprint, amsmath, amssymb, aps, pra, ]{revtex4-2}

\usepackage{graphicx}
\usepackage{upgreek}
\usepackage{dcolumn}
\usepackage{bm}
\usepackage{siunitx}
\usepackage{braket}
\usepackage[dvipsnames]{xcolor}
\usepackage[normalem]{ulem}
\usepackage[utf8]{inputenc}
\usepackage{booktabs} 
\usepackage{makecell}
\usepackage{leftindex}
\usepackage{comment}

\usepackage{xargs}
\usepackage[colorinlistoftodos,prependcaption,textsize=tiny]{todonotes}
\newcommandx{\chris}[2][1=]{\todo[linecolor=OrangeRed,backgroundcolor=OrangeRed!25,bordercolor=red,#1]{#2}}
\newcommandx{\felix}[2][1=]{\todo[linecolor=Blue,backgroundcolor=Blue!25,bordercolor=blue,#1]{#2}}

\newcommand{\imu}{\text{\rm i}}
\newcommand{\diff}{\text{\rm d}}
\newcommand{\expu}{\text{\rm e}}

\newcommand{\spindown}[1]{\hat\sigma_{#1}}
\newcommand{\spinup}[1]{\spindown{#1}^\dagger}

\newcommand{\fieldmode}[2]{\hat{#1}_{#2}}

\newcommand{\fieldmodeatom}[1]{\fieldmode{b}{#1}}
\newcommand{\inmodeatom}[1]{\fieldmodeatom{#1}^\text{in}}

\newcommand{\wgmode}[1]{\fieldmode{a}{#1}}

\newcommand{\fieldmodevac}[1]{\fieldmode{v}{#1}}
\newcommand{\inmodevac}[1]{\fieldmodevac{#1}^\text{in}}

\newcommand{\totalAngularMomentum}{{\bf \hat S}^2}

\newcommand{\betab}{\beta^-}
\newcommand{\betaf}{\beta^+}

\newcommand{\betaUni}{\beta}

\newcommand{\ham}{\hat{H}}
\newcommand{\density}{\hat {\rho}}
\newcommand{\tr}{\text{Tr}}
\newcommand{\OmBold}{{\bf \Omega}}
\newcommand{\kernel}{{\bf \hat \Delta}(\OmBold)}

\newcommand{\ddt}{\frac{d}{dt}}
\newcommand{\ha}{\text{h.c.}}
\newcommand{\cc}{\text{c.c.}}
\renewcommand{\vec}[1]{\boldsymbol{#1}}
\renewcommand{\Re}{\text{Re}}
\renewcommand{\Im}{\text{Im}}

\newcommand{\tlimit}{t_\text{limit}}
\newcommand{\nth}[1]{#1^\text{th}}

\usepackage{hyperref}
\hypersetup{
    colorlinks,
    linkcolor={black},
    citecolor={blue!80!black},
    urlcolor={blue!80!black}
}

\begin{document}


\title{Predicting correlations in superradiant emission from a cascaded quantum system}

\newcommand{\affilHU}{Department of Physics, Humboldt-Universität zu Berlin, 10099 Berlin, Germany}
\newcommand{\affilRPTU}{Department of Physics and Research Center OPTIMAS,
RPTU Kaiserslautern-Landau, 67663 Kaiserslautern, Germany}

\author{Felix Tebbenjohanns}
\affiliation{\affilHU}
\author{Christopher D. Mink}
\affiliation{\affilRPTU}
\author{Constanze Bach}
\affiliation{\affilHU}
\author{Arno Rauschenbeutel}
\affiliation{\affilHU}
\author{Michael Fleischhauer}
\affiliation{\affilRPTU}

\date{\today}

\begin{abstract}
In recent experiments, a novel type of cascaded quantum system has been realized using nanofiber-coupled cold atomic ensembles. This setup has enabled the study of superradiant decay of highly excited collective spin states of up to a thousand atoms, featuring unidirectional coupling mediated by the waveguide mode. 
The complexity arising from the large, multi-excited ensemble and the cascaded interactions between atoms makes conventional simulation methods unsuitable for predicting the correlations of superradiant emission beyond the first order.
To address this challenge, we developed a new simulation technique based on the truncated Wigner approximation for spins. 
Our stochastic simulation tool can predict the second-order quantum coherence function, $g^{(2)}$, along with other correlators of the light field emitted by a strongly excited cascaded system of two-level emitters. This approach thus provides an effective and scalable method for analyzing cascaded quantum systems with large numbers of particles.
\end{abstract}

\maketitle

\section{Introduction}

Simulating the dynamics of a many-body system typically requires computational effort that scales exponentially with the number of particles due to the exponential size of the configuration space. 
Many problems, especially classical ones, can be efficiently approximated using various simulation techniques, such as Monte-Carlo sampling. 
For dynamical quantum many-body problems, however, Monte-Carlo methods can not always be applied.
One significant example of such a system is an ensemble of two-level atoms with dipoles coupled through the quantized electromagnetic field~\cite{hammerer2010quantum}. 
In a typical experimental setup, effective two-level atoms are prepared in a well-defined initial state and probed using resonant or near-resonant light. 
Under certain experimental conditions, the large Hilbert space can be truncated, enabling efficient calculation of the ensemble dynamics. 
For instance, in the weakly driven regime where the atoms are mostly in the ground state, the ensemble behaves as a linear system of harmonic oscillators, which can be solved efficiently~\cite{scully2006directed, eberly2006emission, courteille2010modification, doespiritosanto2020collective}. 
However, even in the few-excitation regime, it remains challenging to predict correlations beyond the Gaussian approximation, such as intensity-intensity correlations of the radiated or scattered field, as a mean-field theory is not sufficient~\cite{kusmierek2023higherorder}. 
Another notable case is a very dense ensemble. 
As Dicke demonstrated, $N$ two-level atoms with pairwise distances much smaller than the wavelength of the emitted light explore only a small number of $N+1$ symmetric Dicke states~\cite{dicke1954coherence, gross1982superradiance}. 
In such cases, realizable in solid-state systems~\cite{cong2016dicke} or, effectively, by using a waveguide or an optical cavity to mediate the coupling between the atoms, a fully inverted ensemble emits its stored energy in a superradiant burst of light~\cite{skribanowitz1973observation, gross1976observation}.

Interestingly, recent theoretical~\cite{cardenaslopez2023manybody} and experimental~\cite{liedl2024observation} studies have revealed that superradiant bursts also occur in \emph{cascaded} quantum systems~\cite{gardiner1993driving, carmichael1993quantum}. 
In such systems, which can be realized through chiral atom-waveguide coupling~\cite{lodahl2017chiral}, information flows unidirectionally through the ensemble. 
Unlike the Hamiltonian studied by Dicke, which is symmetric under particle exchange, the Hamiltonian of a cascaded atomic system lacks this symmetry~\cite{stannigel2012driven, pichler2015quantum, kumlin2020nonexponential}. 
Consequently, the Hilbert space cannot be truncated similarly, making the problem of a driven-dissipative cascaded quantum system beyond the weak drive limit intrinsically exponentially complex.
Current state-of-the-art experiments with chirally coupled two-level systems typically involve either a small number of atoms ($N \ll 10$)~\cite{luxmoore2013interfacing,
soellner2015deterministic,stiesdal2021controlled,joshi2023resonance} or weak coupling to the waveguide~\cite{mitsch2014quantum,liedl2023collective}. 
In the former scenario, numerical tools such as QuTiP can solve the full dynamics. 
In the latter case, the cascaded system is sufficiently coupled to an external reservoir, which allows for a semiclassical description that accounts for leading-order quantum effects while being numerically inexpensive even for a large number of particles.
For example, some of the authors have recently introduced a model with linear computational complexity which quantitatively predicts the intensity radiated by an atomic ensemble that is weakly chirally coupled to a waveguide~\cite{liedl2024observation}. 
However, this model cannot compute higher-order correlations, even in the case of weak coupling. Notably, this includes the second-order quantum coherence function, $g^{(2)}$, which has been explored in a recent experiment measuring the intensity-intensity correlations of a superradiant burst of light~\cite{bach2024}.
\begin{figure}
    \centering
    \includegraphics[width=0.9\columnwidth]{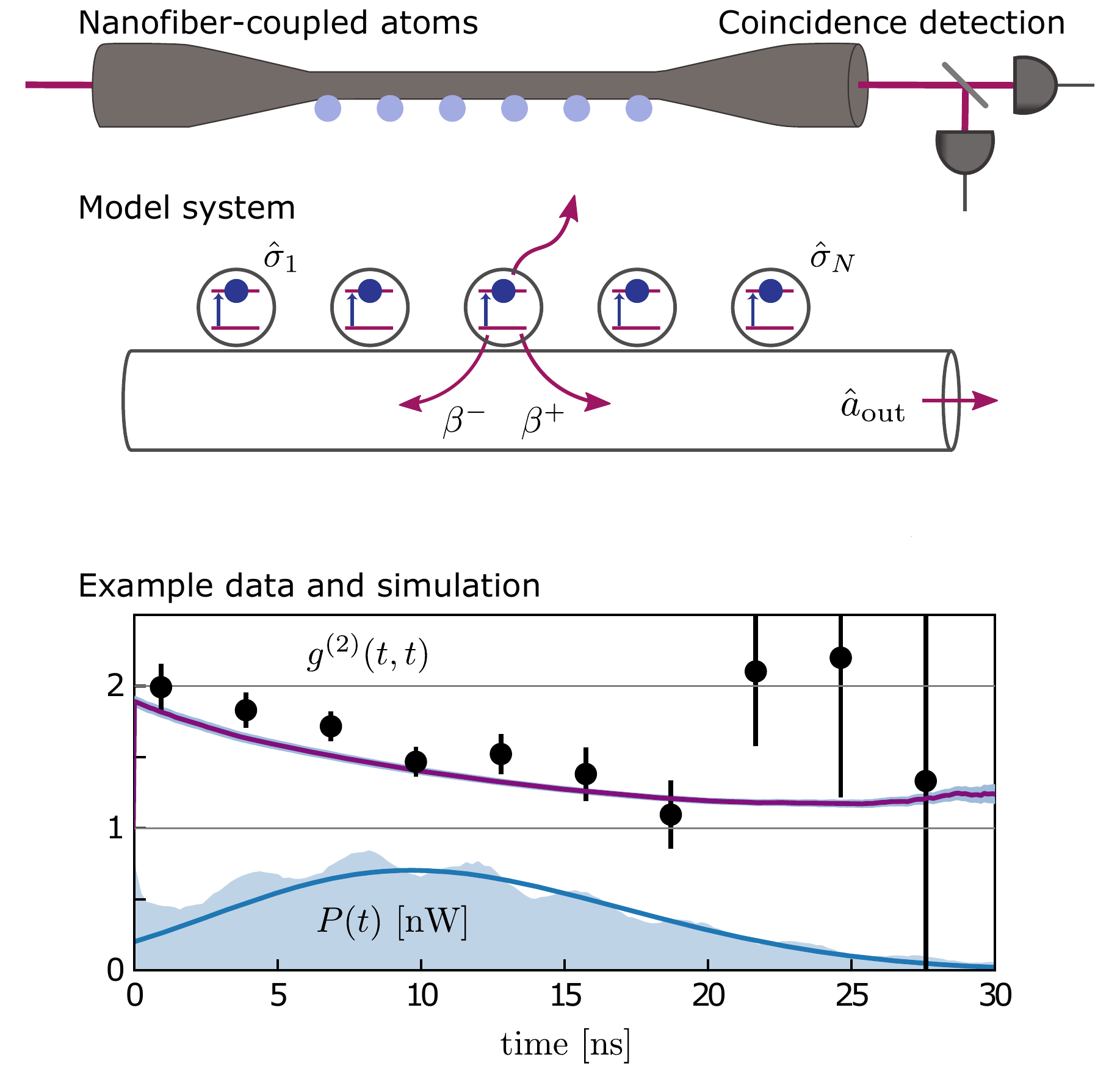}
    \caption{
    Schematic of the setup. $N$ two-level atoms are coupled to a waveguide. 
    The coupling constant to the waveguided mode is direction-dependent, and we assume $\beta^+ \gg \beta^-$, where $\beta^+$ and $\beta^-$ determine the coupling to the forward and backward direction, respectively. This realizes a cascaded quantum system.
    We are interested in correlators of the output mode $\wgmode{\text{out}}$, such as the output power $P(t) = \braket{\wgmode{\text{out}}^\dagger \wgmode{\text{out}}}$ and Glauber's second-order quantum correlation function $G^{(2)}(t,t) = \braket{\wgmode{\text{out}}^\dagger \wgmode{\text{out}}^\dagger  \wgmode{\text{out}}\wgmode{\text{out}}}$.
    In Ref.~\cite{bach2024}, this model is implemented using the D2 transition of nanofiber-coupled cold cesium atoms (excited state lifetime 30~ns). In the bottom panel, we show experimental data from Ref.~\cite{bach2024} of $P(t)$ (blue shaded area) and $g^{(2)}(t,t) = G^{(2)}(t,t) / P(t)^2$ (black data points) for the ensemble decaying from the fully inverted state $\ket{e\cdots e}$. 
    The blue and purple solid lines are the corresponding theoretical predictions, which we calculated using a truncated Wigner approximation approach, as laid out in this work.
    }
    \label{fig:setup}
\end{figure}
Here, we apply another semiclassical model to a cascaded system of weakly coupled quantum emitters subject to drive and losses.  This model has recently been put forward by some of the authors~\cite{mink2022hybrid,mink2023collective} and extends the discrete truncated Wigner approximation for spins~\cite{schachenmayer2015manybody}.

The manuscript is structured as follows.
In Sec.~\ref{sec:master_equation}, we review the master equation for an ensemble of waveguide-coupled two-level atoms, focusing on the cascaded system.
We discuss its relation to the dissipative Dicke model and show that a cascaded quantum system cannot be reduced to the latter.
In Sec.~\ref{sec:HeisenbergLangevin}, we model the system using a set of quantum Langevin equations in the Heisenberg picture, which assume a particularly simple form. 
In our main section~\ref{sec:TWA}, following a brief review of the truncated
Wigner approximation for spins, we derive the stochastic differential equations describing the cascaded system.
In Sec.~\ref{sec:higher_order_correlators}, we present an efficient method to compute relevant correlators of the field radiated by the cascaded system.
Specifically, we calculate the time-dependent power, intensity-intensity correlation, and total angular momentum for up to one thousand atoms, which are initialized in the fully inverted state. 
Beyond its general significance, our model has been successfully employed for the theoretical analysis of the measurement results presented in the aforementioned experimental work~\cite{bach2024}. In Fig.~\ref{fig:setup}, we present a sample set of experimental data along with its theoretical prediction.

\section{Waveguide QED: master equation}\label{sec:master_equation}

The system under consideration is sketched in the middle panel of Fig.~\ref{fig:setup}. An ensemble of $N$ two-level atoms with energy spacing $\omega_{eg}$ is coupled to the modes of a single-mode waveguide.
We label the atoms with indices $n=1,\dots,N$ in ascending order of their non-overlapping positions $z_1 < z_2 < \dots < z_N$ along the waveguide.
As is commonly considered in the literature~\cite{allen2012optical, Scully_Zubairy_1997, carmichael1993open}, we treat the interaction between the atoms and their surrounding electromagnetic field within dipole and rotating wave approximations, we trace out the light field, and we eliminate it using a Born-Markov approximation.
This results in a Lindblad master equation for just the atomic degrees of freedom (we set $\hbar = 1$)~\cite{calajo2022emergence, sheremet2023waveguide}
\begin{subequations}
\begin{align}
    \ddt \density =& -\imu [\ham, \density] + \sum_{mn} \Gamma_{mn} \left( \spindown{n} \density \spinup{m} - \frac{1}{2} \{\spinup{m} \spindown{n}, \density \} \right),\\
    \ham =& -\sum_n \Delta_n \spinup{n} \spindown{n} + \frac{1}{2}\sum_n \left(\Omega_n \spinup{n} + \Omega_n^* \spindown{n} \right) \nonumber\\
    &+ \sum_{mn} J_{mn} \spinup{m} \spindown{n}. \label{eq:Hamiltonian}
\end{align}
\end{subequations}
Here, $\{A,B\} = AB + BA$ denotes the anticommutator, $\spindown{n} = \ket{g_n} \bra{e_n}$ is the spin-lowering operator of the $\nth{n}$ atom and $\ket{g_n}$ and $\ket{e_n}$ are, respectively, the ground and excited state of the $\nth{n}$ atom.
Further, $\Omega_n$ is the Rabi frequency due to a classical driving field and $\Delta_n$ denotes the detuning between the atoms and the field. 
The radiative interactions $J_{mn} = -\imu (V_{mn} - V_{nm}^*) / 2$ and collective decay rates $\Gamma_{mn} = V_{mn} + V_{nm}^*$ are respectively given as the hermitian and antihermitian part of the matrix $V_{mn}$. 
The latter is given by the matrix element~\cite{sheremet2023waveguide}
\begin{equation}
    V_{mn} \propto \vec{d}_m^* \cdot \tensor{G}(z_m, z_n, \omega_{eg}) \cdot \vec{d}_n,
\end{equation}
where we omitted a real-valued prefactor. Here, $\tensor{G}$ is the electromagnetic Green's tensor in the presence of the waveguide, and $\vec{d}_n$ is the atomic dipole moment of the $\nth{n}$ atom.
If a transversal magnetic field is applied to the atoms, their preferred direction of emission into the waveguide can be controlled by tuning their polarization.
More specifically, linearly polarized light results in a bidirectional emission whereas circularly polarized light leads to unidirectional propagation of light into either direction $\pm z$.
With a closer analysis of the Green's tensor~\cite{dzsotjan2010quantum, dzsotjan2011dipole, lekien2017nanofiber}, the matrix element $V_{mn}$ can be approximated in practice by
\begin{equation} \label{eq:collectiveRates}
    \frac{V_{mn}}{\Gamma_0} = \begin{cases}
        \frac{1}{2}, & m=n\\
        \betaf e^{\imu k_z z_{mn}}, &m>n\\
        \betab e^{-\imu k_z z_{mn}}, &m<n
    \end{cases}, 
\end{equation}
where $k_z = n_\text{eff}\omega_{eg} / c$ is the wavenumber of the guided mode with effective refractive index $n_\text{eff}$, and $z_{mn} = z_m - z_n$ is the signed distance between atoms $m$ and $n$.
Furthermore, $\Gamma_0$ is the inverse life time of the excited state of a single atom coupled to the waveguide, and $\betaf$ and $\betab$ are the coupling strengths of an atom to the forward- and backward-propagating waveguide mode, respectively.
A single excited atom emits a photon into the forward-propagating waveguide mode with probability $\betaf$, into the backward-propagating waveguide mode with probability $\betab$, and into the free space with probability $1-\betaf-\betab$, see Fig.~\ref{fig:setup} for a depiction. 
Here, we assume that the atoms do not interact through free space, thus effectively assuming a separation between atoms that is large enough.
The explicit expressions for the rates are then
\begin{subequations} \label{eq:generalCollectiveRates}
\begin{align}
    \frac{J_{mn}}{\Gamma_0} &= \frac{\text{sgn}(z_{mn})}{2\imu }   \left( \betaf e^{\imu k_z z_{mn}} - \betab e^{-\imu k_z z_{mn}} \right), \label{eq:generalInteractionRate}\\
    \frac{\Gamma_{mn}}{\Gamma_0} &= \begin{cases}
        1, &m=n,\\
        \betaf e^{\imu k_z z_{mn}} + \betab e^{-\imu k_z z_{mn}}, & m\neq n
    \end{cases},
    \label{eq:generalDecayRate}
\end{align}
\end{subequations}
with the signum function sgn$(x)$.
When $\betaf = \betab = 0$, the atoms only independently emit into the free space, whereas $\betaf = 1$, $\betab = 0$ indicates that every photon is emitted into the $+z$-direction of the waveguide.

Let us now turn to the experimentally realized case~\cite{liedl2024observation, bach2024} of partial coupling to a unidirectional waveguide, i.e.~$\betab \approx 0$ and $\betaf < 1$.
To simplify notation, we will omit the superscript $+$ in the following, i.e. $\betaUni = \betaf$.
Additionally, a resonant coherent field with amplitude $\alpha$ propagating through the waveguide can be included by substituting $\Delta_n = 0$ and $\Omega_n = 2\alpha \sqrt{\betaUni\Gamma_0} e^{i k_z z_n}$ into \eqref{eq:Hamiltonian}. The complex-valued field amplitude $\alpha$ is scaled such that $|\alpha|^2$ is the photon flux through the waveguide at the input. 
In the following, we measure time in units of the excited state lifetime, such that $\Gamma_0 = 1$. The master equation of the chiral waveguide thus reads~\cite{lekien2008cooperative, stannigel2012driven, pichler2015quantum, cardenaslopez2023manybody, mahmoodian2020dynamics}
\begin{equation}\label{eq:master}
    \ddt \density = -\imu [\hat H_0 + \hat H_\text{casc}, \density] + \mathcal{L}_\text{coll}[\density] + \mathcal{L}_0[\density]
\end{equation}
with ($\ha$ stands for Hermitian conjugate)
\begin{subequations} \label{eq:cascaded_superoperators}
\begin{align}
    \hat H_0 &=  \sqrt{\betaUni} \sum_n  \left( \alpha \spinup{n} + \ha \right) ,\\
    \hat H_\text{casc} &= -\frac{\imu}{2} \betaUni \sum_m \sum_{n<m} \left( \spinup{m}\spindown{n} - \ha \right), \label{eq:cascadedHamiltonian}\\
    \mathcal{L}_\text{coll}[\density] &= \betaUni \sum_{m,n} \left( \spindown{n} \density \spinup{m} - \frac{1}{2} \left\{ \spinup{m}\spindown{n}, \density \right\} \right),  \label{eq:collectivedecay}\\
    \mathcal{L}_0[\density] &= \left(1 - \betaUni \right) \sum_n \left( \spindown{n} \density \spinup{n} - \frac{1}{2} \left\{ \spinup{n}\spindown{n}, \density \right\} \right).
\end{align}
\end{subequations}
Decay to free space modes is described by the Lindblad term $\mathcal{L}_0[\density]$, while the (cascaded) interaction Hamiltonian $\hat H_\text{casc}$ as well as the collective decay Lindblad term $\mathcal{L}_\text{coll}[\density]$ are responsible for the collective dynamics in this cascaded quantum system.
Here, we applied the transformation $e^{-\imu k_0 z_n}  \spindown{n} \rightarrow \spindown{n}$, which simplifies the expressions and which is equivalent to a co-rotating frame of reference. Note that this is only possible in the case of unidirectional coupling.

The goal of this work is to find numerical predictions for time-dependent correlators of the output of the waveguide,
\begin{equation}\label{eq:output}
    \wgmode{\text{out}} = \alpha - \imu \sqrt{\betaUni} \sum_n \spindown{n},
\end{equation}
such as the field $E(t)=\braket{\wgmode{\text{out}}(t)}$,
the output flux $P(t) = \braket{\wgmode{\text{out}}^\dagger(t) \wgmode{\text{out}}(t)}$, and the second-order correlation $G^{(2)}(t,t) = \braket{\wgmode{\text{out}}^\dagger(t) \wgmode{\text{out}}^\dagger(t) \wgmode{\text{out}}(t) \wgmode{\text{out}}(t)}$.
In principle, one could directly solve the master equation~\eqref{eq:master} and then derive the output correlators. 
In practice however, the numerical solution of Eq.~\eqref{eq:master} is not accessible for $N\gg10$ due to the exponentially large Hilbert space.

Various possibilities to obtain approximate solutions have been put forward. 
Since for weakly driven ensembles, the dynamics only depend on the optical density of the ensemble, which is proportional to the product of the number of atoms $N$ and the coupling constant $\betaUni$, one can reduce the complexity by decreasing $N$ while increasing $\betaUni$. 
While this methods ceases to work in principle for strongly driven ensembles, it has been used with some success in a free-space system in Ref.~\cite{ferioli2024nongaussian}. There, the authors approximate their ensemble consisting of some thousands of atoms with finite distances by a model system of about 10 atoms with perfect particle-exchange symmetry, as envisioned by Dicke~\cite{dicke1954coherence}. This reduces the dynamics to 
$\ddt \density = -\imu \alpha [\hat S + \hat S^\dagger, \density] + \hat S\density \hat S^\dagger - \{\hat S^\dagger \hat S, \density\}/2$, with the totally symmetric lowering operator $\hat S = \sum_n\spindown{n}$, such that
the system stays in a small part of the Hilbert space and the above mentioned correlators can be efficiently computed. 
In a cascaded quantum system, however, even in the absence of free space decay, the cascaded contribution of Eq.~\eqref{eq:cascadedHamiltonian} results in the excitation of less cooperative states and thus impedes a numerically inexpensive treatment along the above mentioned lines.

\section{Heisenberg-Langevin equations}\label{sec:HeisenbergLangevin}

As a first approach, let us express the master equation~\eqref{eq:master} equivalently as a set of quantum Langevin equations for the individual spin operators $\spindown{n}$~\cite{gardiner1985input}. In App.~\ref{app:langevin2master}, we show explicitly that for a cascaded quantum system these take the particularly simple and intuitive form ($n=1,\dots,N$)
\begin{equation}\label{eq:langevin}
    \frac{\diff\spindown{n} }{\diff t} = -\frac{1}{2}\spindown{n} -\imu (1-2\spinup{n}\spindown{n}) \left(\sqrt{\betaUni}\wgmode{n}+ \sqrt{1-\betaUni} \inmodevac{n} \right).
\end{equation}
Here, $\wgmode{n}$ and $\inmodevac{n}$ are field operators of, respectively, the waveguide and the free-space modes impinging on the $\nth{n}$ atom. These fulfil the bosonic commutator relations $[\wgmode{n}(t), (\wgmode{n})^\dagger(t')] = [\inmodevac{n}(t), (\inmodevac{n})^\dagger(t')] = \delta(t-t')$, and $[\wgmode{n}(t), (\inmodevac{n})^\dagger(t')]=0$.
For the first atom in the chain, $\wgmode{1}(t) = \alpha(t)$ is the coherent input field.
Because of the unidirectional waveguide, the field operator $\wgmode{n+1}(t)$, which appears in the quantum Langevin equation for atom $n+1$ as an input mode, is identified with the output mode of atom $n$. This relationship is given by the input-output equation ($n=1,\dots,N$)~\cite{gardiner1985input}
\begin{equation}\label{eq:inputoutput}
    \wgmode{n+1} = \wgmode{n} - \imu \sqrt{\betaUni} \spindown{n}.
\end{equation}

Equations \eqref{eq:langevin} and \eqref{eq:inputoutput} represent a complete description of the cascaded quantum system. Note that the second term of Eq.~\eqref{eq:langevin} is non-linear and is responsible for making the solution to these equations difficult to obtain, in general.
In the limit of weak atomic excitation, one can neglect the term $\spinup{n}\spindown{n}\approx 0$, resulting in linear equations of motion, which can be solved analytically~\cite{doespiritosanto2020collective, pennetta2022collective}.

Here, we assume that the first atom is driven by a coherent state, effectively obtaining the optical Bloch equations for the first atom.
Importantly, we cannot assume any other of the fields $\wgmode{n}$ with $n\geq2$ to be coherent, as the field radiated by a two-level atom is famously not coherent in general.
In Ref.~\cite{liedl2024observation}, some of us presented a heuristic model, where the photonic state of the field $\wgmode{n}$ is approximated by a classical mixture of coherent states with the complex amplitude $\alpha_n(\phi) = \sqrt{P_n^c(t)} + \expu^{\imu \phi} \sqrt{P_n^\text{inc}(t)}$, where the angle $\phi$ is drawn from a uniform distribution on the interval $(0,2\pi)$.
This model, in which the field $\wgmode{n}$ has both a coherent and an incoherent contribution with respective flux $P_n^c(t)$ and $P_n^\text{inc}(t)$, 
accurately predicts low-order correlators such as the output field $E(t)$ and flux $P(t)$ with a linear computational complexity in the number of atoms.
Importantly, this method cannot, by design, predict higher-order correlators, such as $G^{(2)}(t,t)$.
In the following, we will therefore apply a novel approximative model of a cascaded quantum system, based on the truncated Wigner approximation, which is both computationally simple and able to predict higher-order correlators.

\section{Truncated Wigner Approximation}\label{sec:TWA}

In this section, we give a short introduction to the truncated Wigner approximation method for spins. For more details, we refer the reader to Refs.~\cite{mink2022hybrid, mink2023collective}. The main result of this section is given by the stochastic differential equations~\eqref{eq:SDE}, which can be evaluated numerically.
Let us consider a linear transformation from the Hilbert space spanned by $N$ two-level atoms to the Wigner phase space~\cite{brif1999phasespace, Klimov2017generalized}. In particular, the elements of the Hilbert space, which are operators $\hat A$, transform into their corresponding Weyl symbol $W_{\hat A}(\OmBold)$
by expanding $\hat A$ into an overcomplete basis
\begin{equation}
    \kernel = \bigotimes_{n=1}^N 
    \frac{1}{2} \begin{pmatrix}
        1+\sqrt{3}\cos\theta_n & \sqrt{3}\expu^{-\imu \phi_n }\sin\theta_n \\
        \sqrt{3}\expu^{\imu \phi_n }\sin\theta_n & 1-\sqrt{3}\cos\theta_n 
    \end{pmatrix}
\end{equation}
with $\OmBold=  \left(\theta_1,\phi_1,\cdots\theta_N,\phi_N\right)$.
This yields
\begin{equation}\label{eq:transformation}
    \hat A = \int \diff \OmBold ~ W_{\hat A}(\OmBold) \kernel,
\end{equation}
where 
\begin{equation}
    \diff \OmBold = \prod_{n=1}^N \frac{\sin(\theta_n)}{2\pi} \diff \theta_n \diff\phi_n.
\end{equation}
The Weyl symbols $W_{\hat A}(\OmBold)$ are elements of the Wigner phase space and can be represented as complex-valued functions on $N$ spheres, i.e.~$\OmBold  \to W_{\hat A}(\OmBold) \in \mathbb{ C}$.

Note that the kernel $\kernel$ is given by a superposition of the spherical harmonics $Y_l^m(\theta_n,\phi_n)$ with $l=0,1$.
Since the spherical harmonics are orthogonal, all spherical harmonics with $l\geq2$ in $W_{\hat A}(\OmBold)$ map to the zero operator $\hat 0$. 
Therefore, the Weyl symbol of some operator $\hat A$ is not uniquely defined, i.e. it possesses a gauge freedom.
Specifically, one can add any spherical harmonic with $l\geq 2$ to a Weyl symbol, without changing the operator it maps to~\cite{mink2022hybrid}.
The Weyl symbol corrresponding to $\hat A$, which only consists of $l=0,1$ spherical harmonics is given by
\begin{equation}\label{eq:op2Weyl}
    W_{\hat A}(\OmBold) = \tr \left[ \hat A \kernel \right].
\end{equation}
For example, the single-atom operators $1_n, \spindown{n}$, and $\spinup{n}\spindown{n}$ are transformed to their corresponding Weyl symbols as
\begin{subequations}
\begin{align}
    W_{1_n}(\OmBold) & = 1, \\
    W_{\spindown{n}}(\OmBold) &= \frac{\sqrt{3}}{2}\expu^{-\imu \phi_n} \sin (\theta_n), \\
    W_{\spinup{n}\spindown{n}}(\OmBold) &= \frac{1+\sqrt{3}\cos(\theta_n)}{2}.
\end{align}
\end{subequations}
Furthermore, for two single-atom operators $\hat A_n$ and $\hat A_m$, which act on different atoms $n\neq m$, we have 
\begin{equation}\label{eq:WeylRuleTwoAtoms}
    W_{\hat A_n \hat A_m}(\OmBold) = W_{\hat A_n}(\OmBold)W_{\hat A_m}(\OmBold),
\end{equation}
while such a factorization is in general not true for two arbitrary operators.

Let us now define the Wigner function $W(\OmBold)$ as the Weyl symbol of the density operator  $\density$, such that 
\begin{equation}\label{eq:wigner}
    \density = \int \diff \OmBold ~ W(\OmBold) \kernel.
\end{equation}
Taking the trace of Eq.~\eqref{eq:wigner} and considering the normalization conditions of density matrix and kernel, $\tr[\density]=\tr\left[\kernel\right]=1$, yields $\int \diff \OmBold~ W(\OmBold) =  1$. 
From the hermiticity of $\density$ and $\kernel$, it follows that $W(\OmBold)\in\mathbb{R}$.
Thus, $W(\OmBold)$ is a quasi probability function. The difference to a proper probability density function (PDF) lies in the fact that $W(\OmBold)$ can be negative for some values of $\OmBold$.

Now, one can map the master equation~\eqref{eq:master} to the Wigner phase space, where it takes the form
\begin{equation}\label{eq:wigner_eom}
    \frac{\partial}{\partial t} W(\OmBold,t) = \mathcal{D} W(\OmBold,t),
\end{equation}
with some high-order differential operator $\mathcal{D}$.
This is achieved by using the so called correspondence rules or Bopp operators which translate the action of an operator on the density matrix $\density(t)$ to a differential operator acting on the Wigner function $W(\OmBold, t)$~\cite{Klimov2017generalized}.
Their exact form for the SU(2) spin operators was first derived in Ref.~\cite{Zueco2006Bopp}.
Since the transformation \eqref{eq:transformation} is linear, we can find the individual differential operators for all terms on the right-hand side of Eq.~\eqref{eq:master} before adding them up.
So far, this transformation of the master equation is exact, and thus finding a solution $W(\OmBold,t)$ is as hard as finding the solution of $\density(t)$.
An approximate solution of $W(\OmBold,t)$ can, however, be obtained efficiently if the following two conditions are fulfilled.
First, we require that at some point in time $t=0$, $W(\OmBold, 0)$ is positive-semidefinite on the whole domain, such that it can be interpreted as a proper PDF.
In fact, it is sufficient if $W(\OmBold, 0)$ is positive after some gauge transformation.
Second, we approximate the differential operator $\mathcal{D}\approx \mathcal{D}_\text{FP}$, where the operator $\mathcal{D}_\text{FP}$ only consists of first and second derivatives with respect to $\theta_n$ and $\phi_n$ in such a way that Eq.~\eqref{eq:wigner_eom} becomes a Fokker-Planck equation~\cite{risken1996fokkerplanck}.
The explicit truncated correspondence rules for many-spin systems are derived in Refs.~\cite{mink2022hybrid,mink2023collective}.
Note that there are different truncation approximations depending on the specific application. 
Different from phase-space approximations to bosonic fields, where the inverse occupation number of relevant modes serves as an expansion parameter, there is not such a general smallness parameter here and a case-to-case discussion of the relevance of the neglected terms is needed.

If both of these requirements are fulfilled, one can first generate a set of angles $\OmBold$ according to the PDF $W(\OmBold,0)$, and then propagate these in time according to a set of coupled stochastic differential equations (SDE) of the form
\begin{equation}
    \diff \Omega_n = f_n (\OmBold) \diff t + \sum_{m=1}^N G_{nm}(\OmBold) \diff W_m(t).
\end{equation}
Here, $\diff W_m(t)$ are Wiener increments with $\overline{\diff W_m(t)}=0$ and 
$\overline{\diff W_n(t)\diff W_m(t)} = \delta_{mn}\diff t $,  
$f_n(\OmBold)$ are functions describing the drift of the system, and 
$G_{nm}(\OmBold)$ are functions describing the diffusion. The overline denotes stochastic averaging.
Both $f_n(\OmBold)$ and $G_{nm}(\OmBold)$ follow from $\mathcal{D}_\text{FP}$ of the Fokker-Planck equation.
These SDEs are numerically solved to produce a large set of trajectories $\OmBold(t)$.
Finally, in order to find a specific atomic correlator $\braket{\hat A(t)}$ at some time $t$, consider the identity 
\begin{equation}
\begin{split}
    \braket{\hat A(t)} &= \tr[\hat A \density(t) ] = \int \diff \OmBold ~ W(\OmBold,t) \tr \left[ \hat A \kernel \right] \\
    &= \int \diff \OmBold ~ W(\OmBold,t) W_{\hat A}(\OmBold).
\end{split}
\end{equation}
Therefore, $\braket{\hat A(t)}$ is given by the phase space expectation value of the function $W_{\hat A}(\OmBold)$.
This can be directly evaluated on the numerically simulated sample set $\OmBold(t)$, i.e.
\begin{equation}\label{eq:wigner_expected}
    \braket{\hat A(t)} \approx \overline{W_{\hat A}(\OmBold(t))}.
\end{equation}
Let us now detail these steps for our cascaded quantum system. At first, we note that any product state $\density = \bigotimes_n \hat{\rho}_n$ maps to a Wigner function, which factorizes as $W(\OmBold) = \prod_n W_n(\Omega_n)$.
Using the aforementioned gauge freedom, the states $\ket{g_n}$ and $\ket{e_n}$ are represented by the positive Wigner functions
\begin{equation}\label{eq:initial}
    W_n(\Omega_n) = \frac{1}{\sin(\theta_n)} \delta \left( \theta_n - \arccos\left(\frac{\pm 1}{\sqrt{3}} \right)\right),
\end{equation}
where the plus (minus) sign is taken for the excited (ground) state.
Positive Wigner functions corresponding to arbitrary other single-particle states characterized by the Bloch-vector $(u,v,w)$ can be found in App.~\ref{app:initial_state}.
At $t=0$, we thus generate a random sample of vectors $\OmBold$, where the components $\theta_n, \phi_n$ are drawn from the PDF $W_n(\Omega_n)$.
The SDEs corresponding to our master equation are described in Ref.~\cite{mink2023collective} and App.~\ref{app:me2sde} and read with $n=1,\dots,N$:
\begin{subequations}\label{eq:SDE}
\begin{align}
    \diff\theta_n &= f_n^{(0)}\diff t  + \Re \left[f_n^\text{coll} \diff t + g_n^\text{coll} \diff Z\right],  \\
    \diff\phi_n &= g_n^{(0)}\diff W_n  - \cot\theta_n \Im \left[f_n^\text{coll} \diff t + g_n^\text{coll} \diff Z\right].
\end{align}
\end{subequations}
Here, the independent atomic decay is modelled by the terms
\begin{subequations}
\begin{align}
    f_n^{(0)}  &=  (1-\betaUni)\left(\cot\theta_n + \frac{\csc\theta_n}{\sqrt{3}}\right) ,\\
    g_n^{(0)}  &= \sqrt{1-\betaUni}\sqrt{1+2\cot\theta_n\left(\cot\theta_n + \frac{\csc\theta_n}{\sqrt{3}}\right)},
\end{align}
\end{subequations}
together with $N$ independent Wiener increments $\diff W_n$. 
In fact, these terms follow exactly from the Lindblad term $\mathcal{L}_0[\density]$ of the master equation without approximation~\cite{mink2022hybrid}.

The collective terms of the master equation, $\hat H_\text{casc}$ and $\mathcal{L}_\text{coll}[\density]$, give rise to a drift and a diffusion term. These terms have been approximated as indicated above, reading
\begin{subequations}
\begin{align}
    f_n^\text{coll} &= \frac{\betaUni}{2}\left(\cot\theta_n + \sqrt{3} \sin\theta_n \right) + 2\imu\sqrt{\betaUni} \expu^{\imu \phi_n} W_{\hat a_n}(\OmBold) ,\\
    g_n^\text{coll} &=-\sqrt{\betaUni}  \expu^{\imu \phi_n},
\end{align}
\end{subequations}
together with the complex-valued Wiener increment $\diff Z$, for which $\overline{\diff Z^2}=0$, $\overline{|\diff Z|^2} = 2\diff t$.
Importantly, $\diff Z$ is the same Wiener increment for all atoms $n$ as it describes the collective coupling to a single guided mode.
Because of the cascaded interaction, $f_n^\text{coll}$ only depends on $W_{\hat a_n}(\OmBold)$, the Weyl-symbol of the field before the $\nth{n}$ atom. This can be solved iteratively according to
\begin{equation}\label{eq:input_output_Weyl}
    W_{\wgmode{n+1}}(\OmBold) = \begin{cases}
        W_{\hat a_{n}}(\OmBold) - \imu \sqrt{\betaUni} W_{\spindown{n}}(\Omega_{n}), & n\geq1 \\
        \alpha, & n=0
    \end{cases},
\end{equation}
which follows directly from the input-output equation~\eqref{eq:inputoutput} and substantially simplifies the calculations.

Summarizing the working principle of this stochastic simulation, we first generate a random set of vectors $\OmBold$ according to the PDF~\eqref{eq:initial}, which represents the initial product state of the atoms. These vectors are then propagated in time according to Eqs.~\eqref{eq:SDE} to Eq.~\eqref{eq:input_output_Weyl}. 
Notice that the number of computations required for accommodating one additional atom is constant, rendering the simulation time a linear function of $N$.
Once the simulation is performed, one obtains the output field directly as
\begin{equation}
    E(t) \approx \overline{W_{\wgmode{N+1}}(\OmBold)}.
\end{equation}
Similarly one could then derive expressions for $P(t)$ and $G^{(2)}(t,t)$. However, a naive implementation of these would lead to a sum over $N^2$ and $N^4$ terms, respectively, spoiling the linear scaling of the computation time. 
In the following we provide an iterative method, which allows the calculation of these (and other) correlators in linear number of computation steps.

\section{Higher-order correlators}\label{sec:higher_order_correlators}

At the core of the cascaded interaction lies the input-output equation~\eqref{eq:inputoutput}, which relates the field operator after the $\nth{n}$ atom to the field operator before it. Based on this equation, we find the following iterative expressions for higher-order correlators of the field after the $\nth{n}$ atom,


\begin{subequations}
\begin{align}
\begin{split}
    \wgmode{n+1}^\dagger\wgmode{n+1} =&~ \wgmode{n}^\dagger\wgmode{n} + \imu \sqrt{\betaUni}\left(\spinup{n}\wgmode{n} - \ha\right) \\
    &~+ \betaUni \spinup{n}\spindown{n}, 
\end{split} \\
\begin{split}
    \wgmode{n+1}^{2\dagger} \wgmode{n+1}^2  =&~ \wgmode{n}^{2\dagger} \wgmode{n}^2 + 2\imu \sqrt{\betaUni} \left(\spinup{n}\wgmode{n}^\dagger \wgmode{n}^2-\ha \right) \\
    & ~ + 4\betaUni \wgmode{n}^\dagger \spinup{n}\spindown{n}\wgmode{n}, 
\end{split} \\
\begin{split}
    \wgmode{n+1}^{\dagger} \wgmode{n+1}^2 =&~ \wgmode{n}^{\dagger} \wgmode{n}^2 - \imu \sqrt{\betaUni} \left(2 \wgmode{n}^{\dagger} \wgmode{n}\spindown{n} - \spinup{n}\wgmode{n}^2 \right) \\
    & ~ +2\betaUni \spinup{n}\spindown{n}\wgmode{n}, 
\end{split} \\
\begin{split}
    \wgmode{n+1}^2 =&~ \wgmode{n}^2 - 2\imu \sqrt{\betaUni} \spindown{n}\wgmode{n},
\end{split} 
\end{align}
\end{subequations}
where we made use of $\spindown{n}^2=0$ and $[\spindown{n}, \wgmode{n}]=0$~\cite{gardiner1985input}.
Note that since the waveguide operator $\wgmode{n}=\alpha - \imu \sqrt{\betaUni}\sum_{k=1}^{n-1} \spindown{k}$ does not act on the $\nth{n}$ atom, Eq.~\eqref{eq:WeylRuleTwoAtoms} yields $W_{\spindown{n} \wgmode{n}} = W_{\spindown{n}}W_{\wgmode{n}}$. With this, the above expressions transform to iterative equations for the corresponding Weyl-symbols as
\begin{subequations}
\begin{align}
\begin{split}
    W_{\wgmode{n+1}^\dagger\wgmode{n+1}} =&~ W_{\wgmode{n}^\dagger\wgmode{n}} + \imu \sqrt{\betaUni}\left(W_{\spindown{n}}^* W_{\wgmode{n}}  - \cc\right) \\
    &~+ \betaUni W_{\spinup{n}\spindown{n}}, 
\end{split} \\
\begin{split}
    W_{\wgmode{n+1}^{2\dagger} \wgmode{n+1}^2}  =&~ 2\imu \sqrt{\betaUni} \left(W_{\spindown{n}}^*W_{\wgmode{n}^\dagger \wgmode{n}^2}-\cc\right) \\
    &~+W_{\wgmode{n}^{2\dagger} \wgmode{n}^2}+ 4\betaUni W_{\wgmode{n}^\dagger \wgmode{n}}  W_{\spinup{n}\spindown{n}} , 
\end{split} \\
\begin{split}
    W_{\wgmode{n+1}^{\dagger} \wgmode{n+1}^2} =&~  - \imu \sqrt{\betaUni} \left(2 W_{\wgmode{n}^{\dagger} \wgmode{n}}W_{\spindown{n}} - W_{\spindown{n}}^*W_{\wgmode{n}^2} \right) \\ 
    &~ +W_{\wgmode{n}^{\dagger} \wgmode{n}^2}+2\betaUni W_{\spinup{n}\spindown{n}}W_{\wgmode{n}}, 
\end{split} \\
\begin{split}
    W_{\wgmode{n+1}^2} =&~ W_{\wgmode{n}^2} - 2\imu \sqrt{\betaUni} W_{\spindown{n}}W_{\wgmode{n}}.
\end{split}
\end{align}
\end{subequations}
For the input field of the first atom, we have $W_{\wgmode{1}^\dagger\wgmode{1}} = |\alpha|^2$,  $W_{\wgmode{1}^{2\dagger}\wgmode{1}^2} = |\alpha|^4$,  $W_{\wgmode{1}^{\dagger}\wgmode{1}^2} = |\alpha|^2\alpha $, and $W_{\wgmode{1}^2} = \alpha^2$. 
These iterative expressions demonstrate that one can compute the Weyl-symbols $W_{\wgmode{n}^\dagger\wgmode{n}}(\OmBold)$ and $W_{\wgmode{n+1}^{2\dagger} \wgmode{n+1}^2}(\OmBold)$ in linear time-complexity, since for each atom a constant number of computations is added.
Finally, using Eq.~\eqref{eq:wigner_expected}, we approximate the optical flux and second-order correlation as
\begin{subequations}
\begin{align}
    P_n(t) \approx \overline{W_{\wgmode{n}^\dagger\wgmode{n}}(\OmBold(t))}, \\
    G^{(2)}_{n}(t,t) \approx \overline{W_{\wgmode{n}^{2\dagger}\wgmode{n}^2}(\OmBold(t))}.
    \intertext{The normalized second-order coherence function can then be computed as}
    g_n^{(2)}(t,t) = \frac{G^{(2)}_{n}(t,t)}{P_n(t)^2} \approx \frac{\overline{W_{\wgmode{n}^{2\dagger}\wgmode{n}^2}(\OmBold(t))}}{\overline{W_{\wgmode{n}^\dagger\wgmode{n}}(\OmBold(t))}^2}.
\end{align}
\end{subequations}
The above arguments can be extended to the calculations of other single-time correlators.

\begin{figure*}
    \centering
    \includegraphics[width=0.9\textwidth]{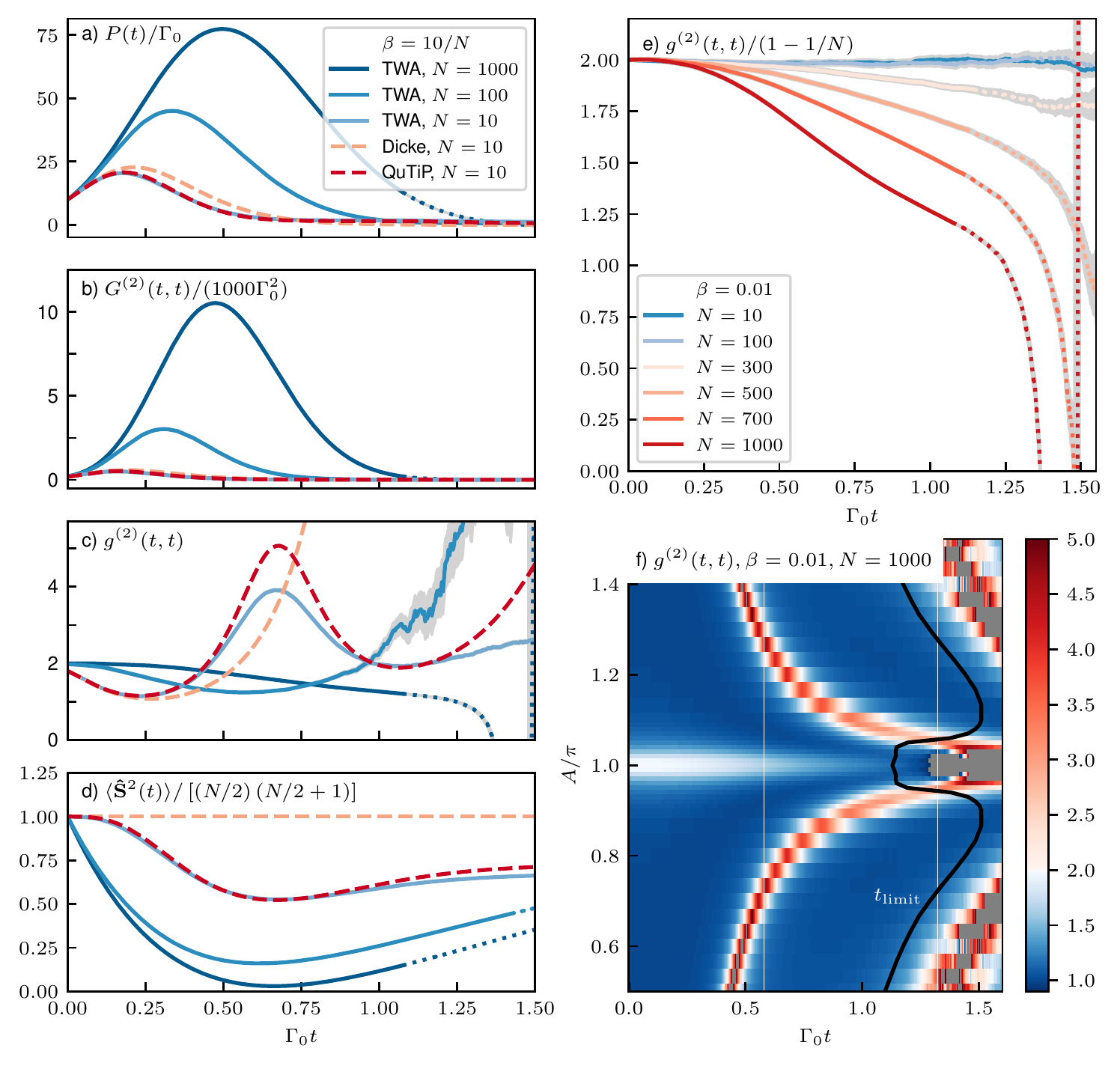}
    \caption{Dynamic simulation of the correlators $P(t)$, $G^{(2)}(t,t)$, $g^{(2)}(t,t)$, and $\braket{\totalAngularMomentum(t)}$ of a cascaded quantum system as indicated.
    In the left panels a) to d), we have $\beta = 10/N$, such that the product $\beta N=10$ is fixed. The initial state is the fully inverted state $\ket{e\cdots e}$, i.e. $A=\pi$ in Eq.~\eqref{eq:initialStateA}.
    In blue, TWA predictions from $N=10, \beta = 1$ (light blue) to $N=1000, \beta = 0.01$ (dark blue). 
    TWA predictions are shown as solid lines for $t<\tlimit$ and as dotted lines for $t>\tlimit$. The time $\tlimit$ is the time when the remaining radiated energy is less than $N/1000$ photons, see main text.
    For comparison, we show the corresponding predictions of the Dicke model for $N=10$ atoms (dashed light red) and the numerical solution of the full cascaded Master equation using QuTiP for $\beta=1,N=10$ (dashed dark red).
    In panel e), we compare $g^{(2)}(t,t)$ for different atom numbers as indicated with $\beta = 0.01$. For more than $1/\beta = 100$ atoms (red colors), second-order quantum coherence builds up during the emission, indicated by the decrease of $g^{(2)}(t,t)$. For fewer than $1/\beta = 100$ atoms (blue colors), $g^{(2)}(t,t)$ is constant.
    In panel f) we show as a color plot $g^{(2)}(t,t)$ for initial states parameterized by the pulse area $A$ according to Eq.~\eqref{eq:initialStateA}.
    In panels a) to e), the shaded grey areas indicate the (one-sigma) uncertainty due to a finite number of simulated trajectories. The number of trajectories was 278,000 for $\beta=0.01$, 15,000 for $\beta=0.1$, and 595,000 for $\beta=1$. For panel f), we simulated 37,000 trajectories.
    }
    \label{fig:results}
\end{figure*}

In Fig.~\ref{fig:results}, we show some simulations which we performed using the truncated Wigner approximation for spins (TWA). In particular, we compute the output correlators $P(t)$ in panel a), $G^{(2)}(t,t)$  in panel b), and $g^{(2)}(t,t) = G^{(2)}(t,t)/P(t)^2$  in panel c) for an ensemble of $N$ cascaded atoms as explained above.
In panel d), we compute the total angular momentum
\begin{equation}
\begin{split}
    \braket{\totalAngularMomentum} &= \braket{\hat S_x^2+ \hat S_y^2+ \hat S_z^2} \\
     &= \frac{1}{4} \sum_{m,n}\braket{\hat\sigma_m^x\hat\sigma_n^x + \hat\sigma_m^y\hat\sigma_n^y + \hat\sigma_m^z\hat\sigma_n^z}
\end{split}
\end{equation} 
in analogy to the other correlators. Here, $\hat\sigma_n^x, \hat\sigma_n^y, \hat\sigma_n^z$ are the Pauli matrices associated with the $\nth{n}$ atom.
Initially, the system is fully excited and we show in blue the TWA simulations of $N=10,100,1000$ atoms with a coupling constant of $\beta = 1,0.1,0.01$, such that the product $\beta N=10$ is constant.
For comparison, we show an exact calculation of the Master equation for $N=10$ atoms using QuTiP (dark red dashed). It can be seen that especially the initial dynamics is well captured by the TWA.
Note that while the TWA's simulation time is linear in $N$, the QuTiP-simulation time is exponential in $N$, making it impossible to compare them to the TWA simulations with $N\gg10$.
Notably, the dynamics of the cascaded quantum system is qualitatively different to that of an ensemble which remains in the symmetric Dicke states~\cite{dicke1954coherence, gross1982superradiance} (light red dashed). This is most clearly visible in panel d), since $\braket{\totalAngularMomentum}$ is maximal and constant for the symmetric Dicke states with $\braket{\totalAngularMomentum} = (N/2)(N/2+1)$~\cite{gross1982superradiance}. In stark contrast, $\braket{\totalAngularMomentum}$ decays in a cascaded quantum system, because the system leaves the symmetrically excited states, even in the loss-less case of $\beta=1$.

We note that, while the TWA captures the early dynamics well, it can fail at large times. There, some of the predictions become even unphysical yielding, e.g., a negative value of $g^{(2)}(t,t)$.
This happens once the optical power in the waveguide is close to vacuum. 
In this limit, the truncation approximation for collective processes used in the TWA simulations becomes inaccurate~\cite{mink2023collective}.
For this reason, we define the time $\tlimit$ by
\begin{equation}
    \int_{\tlimit}^\infty\diff t ~ P(t) = \frac{N}{1000},
\end{equation}
i.e., the integrated flux after $\tlimit$ is equivalent to less than $N/1000$ photons. For $t>\tlimit$, we expect a large systematic error on the TWA predictions, and indicate this by the dotted lines in panels a) to e) and by the black solid line in panel f).
It can be seen that the unphysical predictions only happen for $t>\tlimit$.

In panels e) and f), we show TWA simulations of the second-order coherence
function $g^{(2)}(t,t)$ for a coupling strength of $\beta = 0.01$.
This is motivated by recent experimental results with these parameters~\cite{bach2024}.
In panel e), the initial state is the fully inverted state $\ket{\psi_0} = \ket{e\cdots e}$, for which the initial value of $g^{(2)}(0,0)$ is $2(1-1/N)$, reminiscent of a thermal source~\cite{loudon2000quantum}.
For a small number of atoms ($N<1/\beta=100$, blue lines), $g^{(2)}(t,t)=g^{(2)}(0,0)$  stays constant as a function of time, because the atoms remain in a product state.
However, for a large atom number ($N>1/\beta=100$, red lines), second-order coherence builds up as $g^{(2)}(t,t)$ tends towards unity and the ensemble becomes entangled.
Importantly, this effect, which was measured experimentally in Ref.~\cite{bach2024}, is a consequence of the collective emission process, for which the same threshold of $N \gtrsim 1/\beta=100$ is known, see Ref.~\cite{liedl2024observation}.

Finally, in panel f), we show $g^{(2)}(t,t)$ as a color plot for $N=1000$ atoms and $\beta=0.01$. Here, the initial state of the ensemble is the product state
\begin{equation}\label{eq:initialStateA}
    \ket{\psi_0} = \bigotimes_{n=1}^N \left[\cos \left(\frac{A}{2}\right) \ket{g_n}- \imu \sin\left( \frac{A}{2}\right) \ket{e_n} \right],
\end{equation}
which is characterized by the Rabi-pulse area $A$.
When $A$ is sufficiently different from $\pi$ ($|A-\pi|\gtrsim 2\pi/\sqrt{N}\approx 0.06\pi$), the ensemble has a sizable dipole moment, making the emitted light coherent, i.e. $g^{(2)}(0,0) \approx 1$.
Interestingly, in that regime of superradiance the TWA simulation predicts a sharp peak of $g^{(2)}(t,t)$ for a finite time $t>0$, where the second-order coherence function can reach values much larger than 2. This peak happens well before the time $\tlimit$, i.e., for times where we expect the TWA predictions to be accurate. This effect may be used for realizing a source of highly bunched pulses of light~\cite{jahnke2016giant}.

Let us finally remark that in our simulations we assume the idealized situation of homogeneous atom-waveguide coupling and of exact preparation of the initial state~\eqref{eq:initialStateA}. For the TWA simulations in both Ref.~\cite{bach2024} and Fig.~\ref{fig:setup}, however, we extended our simulations to incorporate experimental nonidealities. Firstly, we simulated the resonant Rabi-pulse with finite pulse length $T\approx 0.13/\Gamma_0$ and pulse area $A$, which is sent through the waveguide~\cite{liedl2024observation}. Secondly, we included inhomogeneous coupling, i.e., each atom features its own coupling strength $\betaUni_n$, which is randomly drawn from an appropriate distribution~\cite{liedl2023collective}. We note that the modelling of these experimental nonidealities is necessary to achieve quantitative agreement between simulation and experiment, and can be readily incorporated into the TWA simulations.

\section{Conclusions}

In this case study, we applied the truncated Wigner approximation for spins to a cascaded quantum system initially prepared in a highly excited state, such as the fully inverted state $\ket{e\cdots e}$. 
We began by reviewing the master equation of the system and the equivalent Heisenberg-Langevin equations, emphasizing that a cascaded quantum system cannot be reduced to the celebrated dissipative Dicke model, even in the limit of a lossless system.
Using the TWA, we demonstrated that the cascaded nature of the system allows for stochastic simulation with a time complexity that scales linearly with the number of atoms in the ensemble. 
We presented simulations of various correlations beyond first order, such as the second-order quantum coherence function $g^{(2)}$ of the output field, for a range of ensemble sizes and coupling strengths. 
To estimate the error introduced by the TWA, we compared its predictions to a full calculation of the master equation, which takes exponential time in $N$, for sufficiently small ensembles. This error is vanishingly small for weak atom-waveguide coupling, $\betaUni \ll 1$, making this simulation method valuable for current experiments involving thousands of nanofiber-coupled atoms~\cite{bach2024}.
In the opposite limit of strong coupling, $\betaUni = 1$, the error remains surprisingly small, indicating that our method is applicable to a broader range of experiments. 
To our knowledge, this is the only computational method in the literature that can handle simulations of intensity-intensity correlations for thousands of atoms in a cascaded quantum system within a feasible timeframe.
Future work will involve testing the applicability of this method for even higher-order correlators, such as three- and four-photon coincidences~\cite{tomm2023photon, drori2023quantum}. 
Additionally, applying the quantum regression theorem to the TWA could enable access to multiple-time correlators like $G^{(2)}(t_1,t_2)$ with $t_1 \neq t_2$, which is an experimentally accessible quantity. 
We expect TWA-based simulations to be useful for many other waveguide QED experiments with large atom numbers, where solving the master equation is infeasible.

\begin{acknowledgments}
We thank Philipp Schneeweiss for insightful discussions.
F.T., C.B., and A.R. gratefully acknowledge funding by the Alexander von Humboldt Foundation in the framework of the Alexander von Humboldt Professorship endowed by the Federal Ministry of Education and Research.
C.M. and M.F. gratefully acknowledge financial support from the DFG through SFB TR 185, project number 277625399.
\end{acknowledgments}

\onecolumngrid
\appendix

\section{Equivalence of Langevin equation approach and master equation.}
\label{app:langevin2master}

In this appendix, we show that for our cascaded quantum system, the combination of the quantum Langevin equations~\eqref{eq:langevin} and the input-output equations~\eqref{eq:inputoutput} are equivalent to the master equation~\eqref{eq:master}.
We mainly apply the formalism put forward by Gardiner and Collett~\cite{gardiner1985input} to a cascaded quantum system. For the special case of two atoms coupled in a cascaded fashion, we refer also to Gardiner's work in Ref.~\cite{gardiner1993driving}.
First, let us consider an arbitrary atomic operator $\hat Q_n$ of a single atom $n$, which can be written as
\begin{equation}
    \hat Q_n = c_0 + c_1 \spindown{n} + c_2 \spinup{n} + c_3\spinup{n}\spindown{n}
\end{equation}
with arbitrary complex numbers $c_0,c_1,c_2,c_3$.
From Eq.~\eqref{eq:langevin}, we find the quantum Langevin equation of $\hat Q_n$ as
\begin{equation}
\begin{split}
     \frac{\diff}{\diff t} & \hat Q_n = c_1 \frac{\diff\spindown{n}}{\diff t} + c_2 \frac{\diff\spinup{n}}{\diff t} +  c_3\left( \spinup{n}\frac{\diff\spindown{n}}{\diff t} + \ha\right) 
    =  
     - \left[ [\hat Q_n,  \spinup{n}] \left(\frac{1}{2} \spindown{n} + \imu \inmodeatom{n} \right)  +  \left(\frac{1}{2}\spinup{n} - \imu (\inmodeatom{n})^\dagger \right)  [\spindown{n},\hat Q_n] \right],
\end{split}
\end{equation}
where $\inmodeatom{n} = \sqrt{\beta_n}\wgmode{n} + \sqrt{1-\beta_n} \inmodevac{n}$ is the input mode of the atom.
Compared to the main text, in this appendix we treat the slightly more general case where the coupling stregth of the atoms can be inhomogenous, i.e. each atom has an individual coupling constant $\beta_n$.
Let us now consider an arbitrary atomic operator of the form $\hat A = \bigotimes_n \hat Q_n$, which we rewrite as $\hat A = \hat A_n^L \bigotimes \hat Q_n \bigotimes \hat A_n^R$, where $\hat A_n^L = \bigotimes_{k<n} \hat Q_k$ only acts on atoms with indices $k<n$, while $\hat A_n^R = \bigotimes_{k>n} \hat Q_k$ only acts on atoms with indices $k>n$. 
Note that $\hat A_n^L$ and $\hat A_n^R$ commute with both $\spindown{n}$ and $\spinup{n}$, and that the terms in the paranthesis, $\spindown{n}/2 + \imu \inmodeatom{n}$ and $\spinup{n}/2 - \imu (\inmodeatom{n})^\dagger$ commute with all system operators, including $\hat A_n^L$ and $\hat A_n^R$~\cite{gardiner1985input}.
Using the product rule, the quantum Langevin equation for $\hat A$ thus reads
\begin{equation}
    \frac{\diff}{\diff t} \hat A =
    \sum_n \hat A_n^L \left(\frac{\diff}{\diff t} \hat Q_n \right) \hat A_n^R = 
    \sum_n  \left\{   [\spinup{n}, \hat A] \left(\frac{1}{2} \spindown{n} + \imu \inmodeatom{n} \right)  +  \left(\frac{1}{2}\spinup{n} - \imu (\inmodeatom{n})^\dagger \right)  [\hat A, \spindown{n}] \right\} ,
\end{equation}
By turning from the Heisenberg to the Schr{\"o}dinger picture, we find
\begin{equation}\label{eq:heisen2schroe}
\begin{split}
    \tr & \left[  \dot{\density} \hat A \right] = \frac{\diff}{\diff t} \braket{\hat A} =  \tr \left[ \density   \frac{\diff}{\diff t}\hat A \right]  = \sum_n \tr \left[ \density 
    \left[[\spinup{n}, \hat A] \left(\frac{1}{2} \spindown{n} + \imu \inmodeatom{n} \right)  +  \left(\frac{1}{2}\spinup{n} - \imu (\inmodeatom{n})^\dagger \right)  [\hat A, \spindown{n}]
    \right]
     \right] \\
    & = \sum_n \tr \left[
    \left[
        \spindown{n}\density \spinup{n}  -  \frac{1}{2}\left\{\spinup{n}\spindown{n}, \density\right\} 
        - \imu \left( 
            \spinup{n}\inmodeatom{n}\density   
            -  \inmodeatom{n}\density \spinup{n} 
            - \density(\inmodeatom{n})^\dagger \spindown{n}
            +  \spindown{n}\density(\inmodeatom{n})^\dagger 
        \right)\right] \hat A \right] \\
    & = \sum_n \tr \left[
    \left[
        \spindown{n}\density \spinup{n}  -  \frac{1}{2}\left\{\spinup{n}\spindown{n}, \density\right\} 
        - \imu \sqrt{\beta_n} \left( 
            \spinup{n}\wgmode{n}\density   
            -  \wgmode{n}\density \spinup{n} 
            - \density(\wgmode{n})^\dagger \spindown{n}
            +  \spindown{n}\density(\wgmode{n})^\dagger 
        \right)\right] \hat A \right].
\end{split}
\end{equation}
In the last two steps, we respectively used the cyclic property of the trace and assumed that the free-space input modes are vacuum, i.e. $\inmodevac{n}\density=0$ and thus $\inmodeatom{n}\density = \sqrt{\beta_n}\wgmode{n}\density$.
Now, we make use of the input-output equation~\eqref{eq:inputoutput} and we expand the input field as $\wgmode{n} = \alpha -\imu \sum_{k=1}^{n-1} \sqrt{\beta_k} \spindown{k}$,
where $\alpha$ is the coherent input field.
Since Eq.~\eqref{eq:heisen2schroe} is correct for arbitrary atomic operators $\hat A$, we can find the master equation of the cascaded quantum system as 
\begin{equation}
\begin{split}
    \frac{\diff}{\diff t} \density 
    =&~ \sum_n \left[ \spindown{n}\density \spinup{n}  -  \frac{1}{2}\left\{\spinup{n}\spindown{n}, \density\right\} 
        - \imu \sqrt{\beta_n} \left( 
            \spinup{n}\wgmode{n}\density   
            -  \wgmode{n}\density \spinup{n} 
            - \density(\wgmode{n})^\dagger \spindown{n}
            +  \spindown{n}\density(\wgmode{n})^\dagger 
        \right) \right] \\
    =&~ \sum_n \left[ \spindown{n}\density \spinup{n}  -  \frac{1}{2}\left\{\spinup{n}\spindown{n}, \density\right\} 
        - \imu \sqrt{\beta_n} 
            [\alpha\spinup{n} + \ha,\density] 
        -  \sum_{k=1}^{n-1}\sqrt{\beta_n\beta_k}  \left( 
            \spinup{n}\spindown{k}\density   
            -  \spindown{k}\density \spinup{n} 
            + \ha
        \right) \right]\\
    =&~ -\imu [\hat H_0, \density] + \mathcal{L}_0[\density]+
        \sum_{k,n}\sqrt{\beta_k\beta_n}  
             \spindown{k}\density \spinup{n}
    - \left[ \frac{1}{2}\sum_n\beta_n \left\{ \spinup{n}\spindown{n} ,\density   \right\}
        +  \sum_{k<n}\sqrt{\beta_n\beta_k}  
            \spinup{n}\spindown{k}\density   
        +  \sum_{k>n}\sqrt{\beta_n\beta_k}   \density\spinup{n}\spindown{k}  
        \right] \\
    =&~ -\imu [\hat H_0 + \hat H_\text{casc}, \density] + \mathcal{L}_0[\density]
    + \mathcal{L}_\text{coll}[\density]
\end{split}
\end{equation}
with

\begin{subequations}
\noindent\begin{minipage}{.4\textwidth}
  \begin{align}
    \hat H_0 &= \alpha\sum_n\sqrt{\beta_n}\spinup{n} + \ha ,\\
    \hat H_\text{casc} &= -\frac{\imu}{2}\sum_{k<n}\sqrt{\beta_k\beta_n} \left(\spinup{n}\spindown{k} - \ha\right),
  \end{align}
\end{minipage}
\begin{minipage}{.6\textwidth}
  \begin{align}
    \mathcal{L}_\text{coll}[\density] &= \sum_{k,n} \sqrt{\beta_k\beta_n} \left( \spindown{k} \density \spinup{n} - \frac{1}{2} \left\{ \spinup{n}\spindown{k}, \density \right\} \right),\\
    \mathcal{L}_0[\density] &= \sum_n\left(1-\beta_n\right) \left( \spindown{n} \density \spinup{n} - \frac{1}{2} \left\{ \spinup{n}\spindown{n}, \density \right\} \right).
  \end{align}
\end{minipage}
\end{subequations}
For homogeneous coupling, i.e. $\beta_n=\betaUni$, one regains the master equation~\eqref{eq:master} from the main text.
By performing each step in reverse, one can also turn the master equation into the quantum Langevin equations. This shows that both descriptions are equivalent.

\section{From master equation to stochastic differential equations}\label{app:me2sde}
The derivation of the stochastic differential equations (SDEs) from the master equation is detailed in Ref.~\cite{mink2023collective}.
The rules described therein can almost directly be applied here, with one exception. There, it was assumed that the coherent coupling coefficients $J_{mn}$ between atoms are real valued. 
In our case, the chiral atom-waveguide coupling leads to complex-valued $J_{mn}$.
The modification due to the imaginary part of $J_{mn}$ is as follows: in the SDEs.~(49) of Ref.~\cite{mink2023collective} one needs to replace $J_{mn} \sin(\phi_{mn})$ by $\Im \left(J_{mn} \expu^{\imu \phi_{mn}} \right)$ and $J_{mn} \cos(\phi_{mn})$ by $\Re \left(J_{mn} \expu^{\imu \phi_{mn}} \right)$.
With this modification, the SDEs of the main text, Eqs.~\eqref{eq:SDE}, follow from Ref.~\cite{mink2023collective}.

In the following table, we summarize the transformation of all terms in the cascaded master equation~\eqref{eq:master} to the corresponding terms in the SDE.

\begin{table}[h]
\setlength{\tabcolsep}{5pt} 
\centering
\begin{tabular}{cc}
\toprule
Term in master equation, $\frac{\diff}{\diff t} \density = $ & Term in SDE, $\begin{pmatrix} \diff\theta_n \\ \diff\phi_n \end{pmatrix} = $ \\
\midrule
\(\displaystyle -\imu [\hat H_0, \density] = -\imu [\sum_n \alpha \sqrt{\betaUni_n} \spinup{n} + \ha, \density]  \) &
\(\displaystyle -2\sqrt{\betaUni_n} \begin{pmatrix}
\Im \left[\expu^{\imu \phi_n} \alpha \right] \\ 
\Re \left[ \expu^{\imu \phi_n} \alpha \right]\cot \theta_n
\end{pmatrix}\diff t \)  \\ \\
\(\displaystyle \mathcal{L}_0[\density] =  \sum_n \left(1 - \betaUni_n \right) \left( \spindown{n} \density \spinup{n} - \frac{1}{2} \left\{ \spinup{n}\spindown{n}, \density \right\} \right) \) & 
\(\displaystyle \begin{pmatrix}
(1 - \betaUni_n )\left(\cot\theta_n + \frac{\csc\theta_n}{\sqrt{3}}\right) \diff t \\ 
\sqrt{1 - \betaUni_n }\sqrt{1+2\cot\theta_n\left(\cot\theta_n + \frac{\csc\theta_n}{\sqrt{3}}\right)} \diff W_n 
\end{pmatrix} \) \\ \\
\makecell{\(\displaystyle \mathcal{L}_\text{coll}[\density] =  \sum_{m,n} \Gamma_{mn}^D  \left( \spindown{n} \density \spinup{m} - \frac{1}{2} \left\{ \spinup{m}\spindown{n}, \density \right\} \right)  \) \\
with $\Gamma_{mn}^D = \sqrt{\betaUni_m \betaUni_n}$
}& 
\(\displaystyle \frac{\sqrt{\betaUni_n}}{2}\begin{pmatrix}
\left[\sqrt{\betaUni_n} \cot\theta_n +  \sqrt{3} \sum_m \sqrt{\betaUni_m} \sin(\theta_m) \cos(\phi_{mn})\right] \diff t - 2\Re \left[ \expu^{\imu \phi_n} dZ\right] \\ 
 \sqrt{3} \cot(\theta_n) \sum_m\sqrt{\betaUni_m} \sin(\theta_m) \sin(\phi_{mn}) \diff t + 2\cot(\theta_n)  \Im \left[  \expu^{\imu \phi_n} dZ\right] 
\end{pmatrix} \) \\  \\
\makecell{ \(\displaystyle -\imu [\hat H_\text{casc}, \density] =  -\imu \left[\sum_{m,n} J_{mn} \spinup{m}\spindown{n}, \density \right] \) \\
with \(\displaystyle J_{mn} = \frac{\sqrt{\betaUni_m \betaUni_n}}{2\imu} \text{sgn}(m-n) \) } & 
\(\displaystyle -\sqrt{3}\sum_m \sin(\theta_m) \frac{\sqrt{\betaUni_m \betaUni_n}}{2}\text{sgn}(m-n) \begin{pmatrix}
 \cos(\phi_{mn})   \\ 
   \cot(\theta_n)  \sin(\phi_{mn})
\end{pmatrix} \diff t \) \\
\bottomrule
\end{tabular}
\label{tab:example}
\end{table}
In Ref.~\cite{mink2023collective}, these rules can be found in Eqs. (16), (49), and (51). For the last rule concerning the cascaded-interaction term $\hat H_\text{casc}$, one needs to use the slight modification as laid out above.
The SDEs in the right column can be rephrased iteratively, as shown in the main text in Eqs.~\eqref{eq:SDE} to \eqref{eq:input_output_Weyl}.

\section{Initial state}\label{app:initial_state}
The discussion around Eq.~\eqref{eq:initial} demonstrates a possible positive semi-definite Wigner function $W(\Omega)$ for a single atom polarized to either its ground state $\ket g$ or its excited state $\ket e$.
In this appendix, we provide a positive semi-definite Wigner function $W(\Omega)$ for an arbitrary single-particle state
\begin{equation}
    \density = \frac{1}{2} \left(1+u \hat \sigma_x + v\hat \sigma_y + w \hat \sigma_z\right),
\end{equation}
where $(u,v,w)=(\braket{\hat \sigma_x}, \braket{\hat \sigma_y},\braket{\hat \sigma_z})$ is the Bloch vector and $\hat \sigma_x, \hat \sigma_y,\hat \sigma_z$ are the Pauli matrices.
Since for any product state of the ensemble, the Wigner function factorizes to $W(\OmBold) = \prod_n W_n(\Omega)$, all such product states can be sampled.
In Ref.~\cite{mink2022hybrid} a possible implementation for any single-particle pure state, i.e.~spin-coherent state, is shown, which can be generated from a rotation of the two states $\ket{g}, \ket{e}$. However, since the Wigner function is not uniquely defined, there are other possible implementations. Here, we present a rather simple one, which allows to simulate even mixed states, i.e. states where the length of the Bloch vector is less than one.
We remind the reader, that a simple evaluation of Eq.~\eqref{eq:op2Weyl} for the state $\density$ does yield a corresponding Wigner function, which however is not positive semi-definite, and can thus not serve as a PDF from which one can sample particular values of $\Omega$.

Consider the Wigner function
\begin{equation}
    W(\Omega) = \frac{A}{\sin(\theta)} \delta(\theta-\theta_w)\left(1+\frac{1}{A}\frac{u\cos\phi + v\sin\phi}{\sqrt{3-w^2}}\right)^2
\end{equation}
with $\theta_w = \arccos(w/\sqrt{3})$ and $
    A = \frac{1}{2}\left(1+\sqrt{1-2\frac{u^2+v^2}{3-w^2}}\right)$.
Note that $A$ is real-valued and positive, which follows with $u^2+v^2+w^2\leq 1$ from
\begin{equation*}
    1-2\frac{u^2+v^2}{3-w^2}\geq  1-2\frac{1-w^2}{3-w^2} = \frac{1+w^2}{3-w^2} >0.
\end{equation*}
From this it can be checked straight-forwardly, that $W(\Omega)\geq 0$ is positive semi-definite. Finally we have
\begin{equation}
    \int \diff \Omega ~ W(\Omega) \begin{pmatrix}
        1 \\ W_{\hat \sigma_x}(\Omega) \\ W_{\hat \sigma_y}(\Omega) \\ W_{\hat \sigma_z}(\Omega)
    \end{pmatrix} = 
    \int \diff \Omega ~ W(\Omega) \begin{pmatrix}
        1 \\ \sqrt{3} \sin(\theta) \cos(\phi) \\ \sqrt{3} \sin(\theta) \sin(\phi) \\ \sqrt{3} \cos(\theta)
    \end{pmatrix}
    = \begin{pmatrix}
        1 \\ u \\ v \\ w
    \end{pmatrix},
\end{equation}
which shows that $W(\Omega)$ transforms to an arbitrary single-atom state $\density$ by Eq.~\eqref{eq:wigner}.

\twocolumngrid
\bibliography{bib}

\end{document}